\documentclass[twocolumn,showpacs,pre,superscriptaddress]{revtex4-1}

\usepackage{graphicx}
\usepackage{amssymb}
\usepackage{amsmath}
\usepackage{amsfonts}
\usepackage{bbm}
\usepackage[colorlinks]{hyperref}
\usepackage[utf8]{inputenc}

\begin{document}
\title{ Coarsening in a 1-D system of Orienting Arrowheads: Persistence with $A+B \rightarrow$ 0 }

\author{Mahendra D. Khandkar}
\affiliation{Department of Applied Physics, Pillai College of Engineering, sec 16, New Panvel - 410206, India }

\author{Robin Stinchcombe}
\affiliation{Rudolf Peierls Centre for Theoretical Physics, University of Oxford, 1, Keble Road, Oxford OX1 3NP, UK.} 

\author{Mustansir Barma}
\affiliation{TIFR Centre for Interdisciplinary Sciences, Tata Institute of Fundamental Research, 21, Brundavan Colony, Osman Sagar Road, Hyderabad 500075, India}

%\date{March 30, 2016}
\begin{abstract}

We demonstrate the large scale effects of the interplay between shape and hard core interactions in a system with left- and right-pointing arrowheads ~$\textless ~~ \textgreater$~ on a line, with reorientation dynamics.  This interplay leads to the formation of two types of domain wall, diffusive ($A$) and static ($B$). The correlation length in the equilibrium state diverges exponentially with increasing arrowhead density, with an ordered state of like orientations arising in the limit. In time, the approach to the ordered state is described by a coarsening process governed by the kinetics of domain wall annihilation $A+B\rightarrow 0$, quite different from $A+A \rightarrow 0$ kinetics pertinent to the Glauber-Ising model. The survival probability of a finite set of walls is shown to decay exponentially in time, in contrast to the power law decay known for $A+A \rightarrow 0$. In the thermodynamic limit with a finite density of walls, coarsening as a function of time $t$ is studied by simulation. While the number of walls falls as $t^{-\frac{1}{2}}$, the fraction of persistent arrowheads decays as $t^{-\theta}$ where $\theta$ is close to $\frac{1}{4}$, quite different from the Ising value. The global persistence too has $\theta=\frac{1}{4}$, as follows from a heuristic argument. In a generalization where the $B$ walls diffuse slowly, $\theta$ varies continuously, increasing with increasing diffusion constant. 

\end{abstract}

\maketitle

\section{Introduction}
One-dimensional systems of interacting particles or spins show interesting collective effects when the system approaches an ordered state as the temperature approaches zero \cite{privman_book,redner_book}. The static properties of such systems are dominated by a diverging correlation length, and generally well understood. However, dynamic properties are more varied and intricate. Of particular interest is the way in which domains of ordered phases grow when the system is quenched from a disordered state to an ordered one.  The coarsening dynamics that ensues can often be modelled through the kinetics of domain walls; a well-known example is the Glauber-Ising chain, in which domain walls diffuse and annihilate upon contact, corresponding to the kinetics of the reaction $A+A \rightarrow$ 0 \cite{privman_book,redner_book}.

\begin{figure}[h!]
\includegraphics[height=4.5cm, width=7.0cm, clip=true]{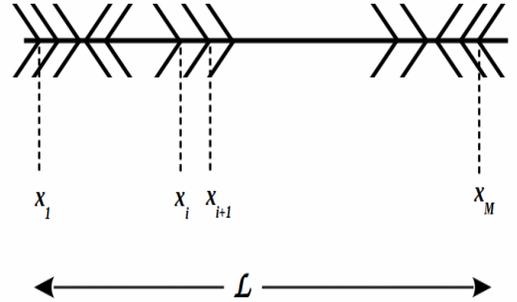}
\caption{Arrowheads in one dimension} 
\label{fig:arrow_heads}
\end{figure}

In this paper we study a system of arrowheads on a continuous 1D line, as depicted in Fig. \ref{fig:arrow_heads}.

Arrowheads resemble bent core or banana-shaped molecules, assemblies of which are known to form ordered phases in higher dimensions \cite{niori,maiti,gurin_1}.  Our objective in studying the simpler one-dimensional problem is to understand, qualitatively and quantitatively, the elements that go into the formation of large stacks of similarly oriented arrowheads. 

These elements turn out to derive from a set of interlinked themes: Entropy-driven ordering; Spin models with assymetric pairwise interaction; Domain wall kinetics with alternating diffusing and stationary walls ($A+B \rightarrow 0$).
As summarized in the discussion below, this sequence leads to novel effects in dynamics, both for a finite number of walls, as well as for a finite wall density in the thermodynamic limit.

Each zero-area arrowhead points right (~$\textgreater$~)  or left (~$\textless$~), and stochastically attempts to change its location (via diffusion) or orientation (by flipping), respecting a no-overlap constraint all the while, as discussed in Section II. Since only hard-core interactions operate (tantamount to the no-overlap constraint) it is evident that purely entropic effects must be responsible for driving the order. In equilibrium, this is brought out explicitly by tracing over locations of arrowheads, thereby generating an effective interaction between successive arrowheads, involving their orientations (Section III). This technique has successfully been used in the past to study 1D assemblies of interacting particles \cite{takahashi, kantor}. In our case, interestingly, this interaction is not symmetric under the interchange of the orientations of a near neighbour pair. A transfer matrix calculation then allows the equation of state and correlation functions to be calculated. Our model is closely related to the `chiral' Ising models of \cite{korea_1,korea_2}, as discussed in more detail in Section II. 

When the degree of alignment is large, there are long stretches of aligned arrowheads, with successive stretches being separated by domain walls. A significant outcome of the non-symmetric interactions is that there are two species of domain walls. In the case of primary interest to us, one species ($A$) diffuses, while the other ($B$) is immobile, with reaction kinetics $A+B \rightarrow 0$. This has a crucial bearing on the dynamics. This is already evident with a finite number of walkers; as shown analytically in Section IV, the full survival probability decays exponentially in time, in strong contrast to the power law decays that characterize $A+A \rightarrow 0$ \cite{fisher}, pertinent to relaxation in the Glauber-Ising model. Moreover, with a thermodynamically large number of walkers, numerical simulations reveal interesting effects (Section V). Further, during the approach to steady state (section VI), the density of domain walls falls as $\sim t^{-\frac{1}{2}}$ as in the Ising case. Persistence properties show greater variation. The persistent fraction of arrowheads is found to decay as a power law $\sim t^{-\theta}$ , with $\theta$ close to $\frac{1}{4}$ , in contrast to the exact value $\frac{3}{8}$ for the Ising case \cite{derrida_2,derrida,satya_3}. The global persistence decays with the same power $\frac{1}{4}$ , a result which can be justified through theoretical arguments, as in the Ising case \cite{satya_1}. 

Finally, on allowing the $B$ walls to diffuse, albeit less slowly than the $A$’s (Section VII), a degree of nonuniversality is revealed: the exponent $\theta$ is found to depend continuously on the ratio of diffusion constants, even though the fraction of domain walls continues to decay as $\sim t^{-\frac{1}{2}}$  . This is reminiscent of the results of \cite{korea_1,korea_2}, where a continuous variation of exponents was found as a function of coupling constants, though in different quantities.

\section{Model}
Our model consists of $M$ arrowheads, each of length $\ell$ on a 1D line of length $L$, with a hard core constraint which implies no overlap (Fig. \ref{fig:arrow_heads}). While $L$ is taken to be of fixed value with periodic boundary conditions while discussing dynamics, it proves expedient to allow $L$ to fluctuate in a constant pressure ensemble in order to derive static properties; in that case, the number density $\rho = \frac{M}{\langle L \rangle}$ and other thermodynamic quantities are well-defined in the thermodynamic limit.

A microscopic configuration $C$ is specified by the set $\{x_i,S_i\}$ where $x_i$ is the coordinate of the vertex of the $i^{th}$ arrowhead while $S_i = \pm 1$ represents the orientation, with the positive sign corresponding to  $\textgreater$ and the negative sign to  $\textless$.

 The full dynamics involves attempts to make either a spin flip or displacement of a randomly chosen arrowhead $i$, as follows \\ 
(i) it can attempt to flip about its vertex $S_i \rightarrow S_i^{\prime} = - S_i$  \\
(ii) it can attempt a displacement $\delta$  where  $\delta$ is chosen with uniform probability in the interval $[x_i - \Delta, x_i + \Delta]$  where $\Delta$ is a fixed length.\\
The attempted moves in (i) and (ii) are accepted only if they do not lead to an overlap with other arrowheads, and do not lead to crossing arrowheads in case (ii). 
As discussed in Section III, these moves respect detailed balance and hence lead the system to an equilibrium state. This state is characterized by a correlation length which diverges as $\rho \rightarrow \infty$, approaching an ordered state with like orientations ($\textless \textless \textless ...$ or $\textgreater \textgreater \textgreater...$). 

Since both arrowhead flip and displacement attempts are involved in the dynamics, the time-dependent properties of the system would be expected to depend on
parameters which govern these, both for autocorrelation functions in steady state, and for the dynamics of approach to steady state. Our preliminary numerical
studies indeed indicate that at finite densities $\rho$, the dynamic power laws do depend on the displacement range $\Delta$. In this paper, however, we are 
primarily interested in the approach to the fully ordered state which is reached only in the limit $\rho \rightarrow \infty$. Accordingly, we will retain only
the re-orientation move (i); the displacement move (ii) is ineffective in this limit. Hence the allowed dynamical moves for the central member in triplet of
successive arrowheads are:
\begin{equation}\label{eq:moves_1}
\begin{split}
\textgreater \textless \textless  & \rightarrow \textgreater \textgreater \textless   ~~~~~ with ~~ rate ~~ u/2 \\
\textgreater \textgreater \textless &  \rightarrow \textgreater \textless \textless   ~~~~~ with ~~ rate ~~ u/2 \\
\textgreater \textless  \textgreater & \rightarrow \textgreater \textgreater \textgreater ~~~~~  with ~~  rate ~~ u \\
\textless  \textgreater \textless & \rightarrow \textless \textless \textless   ~~~~~ with ~~ rate ~~ u             
\end{split}
\end{equation}
Note that the triplets $\textless \textless \textless$, $\textgreater \textgreater \textgreater$, $\textless  \textgreater \textgreater$, $\textless \textless  \textgreater$  cannot evolve in the limit considered, owing to the no-overlap constraint. 

These rules have an important implication for interfaces which separate ordered segments of similarly-oriented arrowheads. Evidently, the interfaces are of two types: ($A$) ...$\textgreater \textgreater \textless \textless$ .... and  ($B$) ....$\textless \textless  \textgreater \textgreater$... . While $A$ interfaces can evolve (and move in the process), $B$ interfaces are static. This distinction is ultimately responsible for the difference of behaviour in the coarsening dynamics vis a vis the Glauber-Ising model, where both types of interface evolve and diffuse at equal rates. 
In order to investigate the effects of allowing $B$ interfaces to diffuse, though more slowly than $A$ interfaces, in Section VI we allow configurations $\textless  \textgreater \textgreater$ and $\textless \textless  \textgreater$  to evolve 
\begin{equation}\label{eq:moves_2}
\begin{split}
\textless  \textgreater \textgreater & \rightarrow \textless \textless  \textgreater ~~~~~ with ~~ rate ~~ u^{\prime}/2 \\
\textless \textless  \textgreater & \rightarrow  \textless  \textgreater \textgreater ~~~~~ with ~~ rate ~~ u^{\prime}/2             
\end{split}
\end{equation}
with $u^{\prime} \textless  u$. Evidently, by varying the ratio $u^{\prime}/u$ between 0 and 1, we generate a family of models which interpolates between arrowhead model and the Ising model.

The dynamical evolution rules are closely related to those considered by Kim et al \cite{korea_1}. These authors studied nonequilibrium Ising models with dynamics which they termed `chiral', namely with different transition rates at ($+,-$) and ($-,+$) kinks. Besides nonconserving single spin flips (analogous to arrowhead reorientations), they allowed ASEP-like moves which conserve spin; these have an important effect on the dynamics, as discussed in subsequent sections.

\section{Equilibrium Static Properties }
\subsection {Introduction }

In this section, we consider the equilibrium properties of the arrowhead model introduced in Section II and shown in Fig. \ref{fig:arrow_heads}. 
Evidently the hard-core constraint between arrowheads plays a crucial role in determining the set of allowed microscopic configurations, i.e.
 the possible arrowhead vertex coordinates and ``spin'' orientations \{$x_i, S_i$\} which specify each allowed configuration $C$.

We first observe that the condition of detailed balance $W(C \rightarrow C^{\prime})Prob(C)= W(C^{\prime} \rightarrow C)Prob(C^{\prime})$ is valid provided Prob($C$) is chosen to be equal for every allowed configuration $C$.  This is because an allowed arrowhead flip $S_i$ $\rightarrow$ $S_i^{\prime}$ occurs at the same rate as the reverse move, and this is true also of every displacement move  $x_i$ $\rightarrow$ $x_i$ + $\Delta$  which leads to an allowed configuration, and its reverse. 
Since every allowed configuration has equal energy, correlations between arrowheads develop purely from entropy. It is well known that in systems with hard core interactions, entropy can lead to a tendency towards ordering \cite{frenkel}. Indeed, as we will see below, this tendency is present in our system as well, and leads ultimately to a diverging correlation length with increasing arrowhead density.

As for several 1D systems with hard objects on a line \cite{kantor,talbot,gurin_2}, our system of arrowheads can be solved exactly by integrating over the coordinates \{$x_i$\}, thereby generating an effective interaction between nearest neighbour spins. The important point in our case is that this interaction is not symmetric under the interchange of spins in a near neighbour pair of Ising spins. As discussed in Section II, our model is closely related to the chiral Ising models discussed in \cite{korea_1,korea_2} which have a similar asymmetry (though no handedness distinction). 
The equilibrium properties of the resulting spin system can be obtained using a transfer matrix technique, resulting in closed form expressions for the equation of state and correlation functions. We also study the extent of spatial persistence in this system.

\subsection {Equation of State }
It is convenient to embody the hard-core constraint between arrowheads by introducing an orientation-dependent potential energy of interaction

\begin{equation}\label{eq:potential}
\begin{split}
V_{++} (x_{i+1}-x_i) & =V_{--} (x_{i+1}-x_i )=V_{+-} (x_{i+1}-x_i )=0      \\
                       & ~~~~~~ \text{if} ~~ (x_{i+1}-x_i ) ~~ \textgreater ~ 0 ,                   \\
                                                                          \\
 V_{-+} (x_{i+1}-x_i) & = \infty ~~~~~~ \text{if} ~~ 0 ~ \textless ~ (x_{i+1}-x_i ) ~ \textless ~ 2\ell \\  
                       & = 0      ~~~~~~~ \text{if} ~~ (x_{i+1}-x_i ) ~ \geq ~ 2\ell          
\end{split}
\end{equation}
 
We work in a constant pressure ensemble, where the pressure $P$ and temperature $T\equiv 1/\beta$ are specified. The coordinate $x_1$ of the first particle is held fixed (though $S_1$ can flip), while all other $x_i$ can fluctuate, implying that the total length of the system $L=(x_M-x_1)$ can fluctuate as well. The corresponding partition function is then      
     
\begin{equation}
\begin{split}
 Q_M  =  & \sum_{\{S_k\}}\prod_{k=2}^M \int dx_k ~ exp [ -\beta \sum_{i=1}^{M-1} \{V_{S_iS_{i+1}} (x_{i+1} - x_i )   \\
          &+ P(x_{i+1} - x_i) \} ]    
\end{split}
\end{equation}

For a specified set of spin orientations {$S_i$ } it is straightforward to perform the integrals over {$x_i$ } sequentially over $i$ with the result

\begin{equation}
     Q_M = \sum_{\{S_i\}} \prod_{i=1}^{M-1} w(S_i,S_{i+1})            
\end{equation}
where $w(+,+)=w(-,-)=w(+,-)=1/\beta P $ ; $ w(-,+)=\frac{exp(-2 \beta P\ell)}{\beta P} $ .  In terms of the transfer operator $W$ with matrix elements $w(S_i,S_{i+1})$, we may write  

\begin{equation}
         Q_M = \sum_{\{S_1,S_M\}} \langle S_1 |\mathbbm{W}^{M-1} |S_M \rangle
\end{equation}
which may readily be evaluated by diagonalizing $\mathbbm{W}$. Let us define $g=exp(-\beta P\ell)$. Then the eigenvalues $\lambda_{\pm}$ are given by ($1\pm g$), with corresponding right eigenvectors  $|e_{\pm} \rangle$ and left eigenvectors $\langle e_{\pm} |$.  The (unnormalized) entries of  $| e_{\pm}\rangle$ are   (1,$\pm g$) while those of $\langle e_{\pm} |$    are ($g, \pm 1$).

\bigskip
 We find

\begin{equation}
         Q_M = (\frac{1}{\beta P})^{(M-1)}   \frac{(1+g)^2}{2g} [(1+g)^{M-1}-(1-g)^{M-1} ]            
\end{equation}

 In the limit of large $M$, we obtain

\begin{equation}
\frac{1}{M}  ln Q_M = ln\frac{1}{\beta P} +ln(1+g)             
\end{equation}

Recalling that the average system length $\langle L \rangle $ = $-\frac {\partial ln Q_M }{\partial \beta P}$  we may find the number density $\rho \equiv \frac{M}{\langle L \rangle} $ as a function of $\beta$  and $P$, yielding the equation of state

\begin{equation}
  \frac{1}{\rho}  =   \frac{1}{\beta P}  +  \frac {\ell}{(1+e^{\beta P\ell})}             
\end{equation}

The contribution to the equation of state coming from the arrowhead configurations is embodied in the second term on the right hand side. This term is a correction to the ideal gas contribution  $1/\beta P$ ; it arises from the hard core interaction between arrowheads. Interestingly, it is small at both high and low values of $\beta P \ell$.  When $\beta P \ell$ is small, arrowheads are well separated and their orientation is unimportant, so that they approximate a free ideal gas. On the other hand, when $\beta P \ell$ is large, the large entropic cost of the pair sequence ($-,+$) makes its occurrence exponentially unlikely. The vertex locations of the remaining sequences of arrowheads are isomorphic to those of an ideal gas of point particles. 
The rare occurrence of ($-,+$), implies that ($+,-$) is equally rare, as these pair sequences must alternate. Together, this implies that the system correlation length must become very large as $T \rightarrow 0$. This is verified by direct calculation as discussed below. 

\subsection {Correlation Function and Spatial Persistence}
The two-point correlation function can be evaluated using the transfer matrix formalism. With free boundary conditions, we have

\begin{equation}
\begin{split}
& C(r)\equiv \langle  S_i S_{i+r} \rangle  =  \\
& Q_M^{-1}\sum_{\{S_1,S_M\}} \langle S_1 | \mathbbm{W}^{i-1} \sigma^z \mathbbm{W}^r \sigma^z \mathbbm{W}^{M-i-r} |S_M\rangle             
\end{split}
\end{equation}
where $\sigma^z$ is the $z$ Pauli matrix. $C(r)$  can be evaluated through a standard route. Assuming that $i$ and $M-i-r$ are both of order $M$,
we obtain 
 $C(r)=(\frac{\lambda_-}{\lambda_+})^r$ = $(\frac{1-g}{1+g})^r$ in the thermodynamic limit $M \rightarrow$ $\infty$. Thus the correlation length $\xi$
 is given by
                                                    $\xi = -1/(ln ~ tanh ~ \frac{\beta P \ell}{2})$.

In the dense packing limit $\beta P \ell \rightarrow \infty$, we obtain $\xi \approx \frac{1}{2}  e^{\beta P \ell} \approx \frac{1}{2}   e^{\rho \ell}$. 

It is also interesting to ask for the spatial persistence, namely the probability that the same value of the spin, say $+1$, occurs unbroken over a stretch of $r$ sites \cite{satya_2}. In terms of the projection operator $n_+$ = $(1+\sigma^z)/2$ with eigenvalues $n_+$=0 or 1 this probability is given by
    $P_{pers} (r) \equiv \langle n_{+,i} n_{+,i+1} .... n_{+, i+r-1} \rangle $  =  $\lambda_+^{-r}\langle e_+ |(\mathbbm{n}_+ \mathbbm{W})^r |e_+ \rangle $
where the thermodynamic limit $M\rightarrow \infty$ has been assumed. Given the form of the operator $\mathbbm{n}_+\mathbbm{W}$, the matrix element on the right hand side is found to be $(1+g)/2$ , independent of $r$. Thus $P_{pers} (r)$ decays exponentially, as $\tilde e^{-r/\zeta}$ where
                                          $\zeta = 1/[ln(1+e^{-\beta Pl})]$.
In the dense packing limit $\beta P \ell \rightarrow \infty$ , we obtain $\zeta \approx  e^{\beta P\ell} \approx e^{\rho \ell} $. 
Thus in the limit of high density, both the correlation length $\xi$  and the persistence length $\zeta$ diverge in similar ways, with prefactors differing by a factor 2. Recall that these `lengths' pertain to spin separations which correspond to arrowhead labels, so they need to be divided by the density in order to convert them to lengths on the line.

Finally we investigate the effect of interaction asymmetry on the spatial separation of arrowheads. Suppose the orientations of the $i^{th}$ and $(i+r)^{th}$ arrowheads have been specified to be $S_i^{\prime}$ and $S_{i+r}^{\prime}$, and their mean separation is $\langle Y(S_i^{\prime},S_{i+r}^{\prime})\rangle$. For a given sequence $\{S_j\}$ of intermediate spins, $Y$ is the sum of separations of successive pairs $y(S_j,S_{j+1})$. From Eq.(\ref{eq:potential}) it is easy to see that $y(+,+)=y(-,-)=y(+,-)=1/\beta P$ while $y(-,+)= \frac{1}{\beta P} + 2\ell$. Thus 

\begin{equation}
\langle Y(S_i^{\prime} ,S_{i+r}^{\prime})\rangle  =  \frac{r}{\beta P} + 2\ell f(S_i^{\prime},S_{i+r}^{\prime})             
\end{equation}

Here  $f(S_i^{\prime},S_{i+r}^{\prime})$ is the mean number of ($-,+$) pairs in the stretch ($i,i+r$); it can be calculated using the transfer matrix within the finite stretch, keeping track of each occurrence of ($-,+$). The result is

\begin{equation}
\begin{split}
f(-,+) & =\frac{1}{2}(1+gr\frac {(1+g)^{r-1}+(1-g)^{r-1}}{(1+g)^r-(1-g)^r} )  \\ \\
f(+,-) & =\frac{1}{2}(-1+gr\frac {(1+g)^{r-1}+(1-g)^{r-1}}{(1+g)^r-(1-g)^r} )            
\end{split}
\end{equation}

The mean numbers of $\textless$ $\textgreater$ pairs differ by exactly 1 for the two specifications of $(S_i^{\prime},S_{i+r}^{\prime})$, implying that 
$\langle Y(-,+) \rangle$    exceeds $\langle Y(+,-) \rangle$ by $2\ell$.
This is a quantitative measure of the effects of asymmetry in the interaction.

\section { Dynamics with `$N$' Walkers}

\subsection { Domain walls, Survival probability}

In the one-dimensional model being considered any possible configuration is a sequence of domains each with all arrows in one direction (~$\textless$ or $\textgreater$~) bounded by domain walls  $\textgreater$$\textless$  or  $\textless$ $\textgreater$.  The dynamic moves are flips of arrows about their vertex with  no overlap allowed between arrowheads. Allowed arrowhead flips like  $\textgreater$$\textgreater$$\textless$  going to  $\textgreater$$\textless$$\textless$  can generate hopping of the domain walls $\textgreater$ $\textless$ (hereafter called ``walkers'' or mobile $A$ - particles  in this Section).    Here we take the limit $\rho$ $\rightarrow$ $\infty$ (corresponding to $T \rightarrow 0$ in the spin model) which makes any (isolated) one of the other type (``walls'' or  $B$ - particles) immobile.

Further, since  $\textless$$\textgreater$$\textless$  can go to  $\textless$$\textless$$\textless$  (etc) we have a process which is a special form of  $A + B \rightarrow 0 $ in which a ``walker'' pair-annihilates with an otherwise fixed ``wall''. The no-overlap constraint prevents the annihilation of pairs of like particles. 

Furthermore, the two types of domain wall/particle, $A$ and $B$, alternate, and continue to do so even after any allowed pair annihilation. But the surviving $B$'s are fixed.  It will be seen that the general system dynamics is qualitatively different from that of $A + A \rightarrow 0$.

In this section we investigate the survival probability  for the general case of $N+1$ ``walls'' with $N$ intervening ``walkers''.
The specific aim is to obtain the probability $Q^{N,N+1}$ (\{$x_k$\}, \{$a_k$\} ; \{$b_j$\}) of survival to time $t$ of all $N$ walkers $k=1$ to $N$, (with initial positions {$a_k$}) to positions  {$x_k$}, and of all walls $j=1$ to  $N+1$ (fixed at {$b_j$}).

This can be treated by an image method of the type used for vicious walkers and related systems \cite{fisher} despite the very different dynamics resulting.

\subsection {Image Method}

\begin{figure}[h!]
\includegraphics[height=4.5cm,width=7.8cm, clip=true]{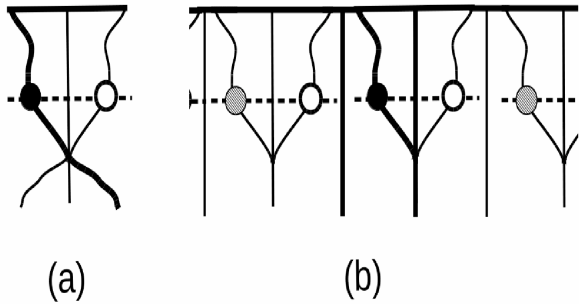}
\caption{Schematics for walkers}
\label{fig:walkers}
\end{figure}

Figure \ref{fig:walkers} (a)  gives the simplest illustration of the image method, for a single walker.

In the figure the thick black line can represent a particular realization of the path of a walker shown by the black circle (at its position $x$ at time $t$, increasing to the right and downwards, respectively). Effects of an added fixed reflecting or absorbing wall, shown by the vertical straight line, can be represented by an image walker, the open circle, together with its path, a reflection (in the wall) of the original path.

This scheme is appropriate for walls of the type $\textless ~ \textgreater$ in the system being studied, which pair annihilate with the walker.
For a single random walker with no wall the ``free'' probability function (over all possible paths starting at $x=a$  at $t=0$) is

\begin{equation}
  \phi(x,a) = (4 \pi t)^{-1/2} exp [ -(x-a)^2 /4t]             
\end{equation}
(taking diffusion constant unity).

The corresponding probability for an image resulting from a wall at $b$ in the manner just described, is  $R_b$ $\phi(x,a) $= $\phi(x,2b-a)$.
The operation $R_b$ corresponds to a sort of reflection in the wall at $x=b$. Then  ($1 - R_b$)$\phi$ vanishes at $x=b$ and is, up to that point,  the appropriate probability for obtaining survival properties for the system of the single walker with pair-annihilating wall at $x=b$.

Similarly for a single walker between two such walls at $b_1$, $b_2$ an image method provides the survival probability distribution function $Q^{1,2} (x,a ; b_1,b_2$).  However here the character of the procedure (after the first stages) and the consequences are very different from the trivial case just described, and representative of those for the general case.

To satisfy the necessary boundary conditions (vanishing of $Q^{1,2}$  at the walls) repeated application of the operations ($1-R_{b_1}$),  ($1-R_{b_2}$) to the free distribution function  $\phi (x,a)$ is required. This gives rise to a proliferation of images whose positions are periodically related at any time, including $t=0$.

Figure \ref{fig:walkers} (b) shows, for a particular realisation, the paths, and positions for a specific time $t$, of the walker and of its images produced in this way.

Here the black circle is the walker, and the dashed and open circles are respectively its positive and negative images (having contributions to $Q^{1,2}$ of opposite signs, coming from the negative sign in the factors ($1-R_{b_1}$),($1-R_{b_2}$). Consequently the survival probability for the $N=1$ case being considered is

\begin{equation}\label{eq:Q12_1}
\begin{split}
& Q^{1,2} (x,a;b_1,b_2)  = \\
                      & \sum_{n=-\infty}^{\infty} [\phi(x+2(b_2-b_1)n,a)-\phi(2b_1 - x + 2(b_2-b_1) n, a)].               
\end{split}
\end{equation}

Any distribution  $\mu(a)$ of initial positions $a$, which generalises the free distribution function  $\phi (x,a)$  to  $\Phi(x;\mu) = \int da ~ \mu(a) ~ \phi(x,a)$ gets similarly proliferated, and superposition allows us to include that in what follows, by replacing $\phi$ in Eq. (\ref{eq:Q12_1}) by  $\Phi$  .

Our discussion so far has not made apparent a crucially important property of the survival probability for the case being considered, namely its exponential time decay. This property, shared with generalisations (below) to unlimited numbers of walkers, is a consequence of the walk's confinement.
The exponential decay is most easily quantified using the Fourier representation of $Q^{1,2}$. The fourier transform of $\phi(x,a)$ is $e^{-q^2t} C_q$ where $C_q \propto e^{-iqa}$. 

It is easy to show that the non-vanishing fourier components of  $Q^{1,2}$ have discrete wave-vectors $q = \{q_n\} = \{\frac{n~\pi}{b_2-b_1}\}$ where $n$ is any non-zero integer (corresponding to the period of the image structure, allowing for its alternation).

The resulting form for    $Q^{1,2}$    is 

\begin{equation}\label{eq:Q12_2}
\begin{split}
& Q^{1,2} (x,a ; b_1,b_2)  =    \\ 
                        & \sum_{n=1}^{\infty} 2 e^{{-q_n}^{2} t} [cos q_n (a-x) - cos q_n (x+a-2b_1)]            
\end{split}
\end{equation}

So at long times  $t \gtrsim$ ${q_1}^{-2}$ $\equiv$ $\tau$  the survival probability decays exponentially $\propto exp(-t/\tau)$        
with decay time  $\tau = (\frac{b_2 - b_1}{\pi})^2$. This $\tau$ applies also for the generalised case having distribution  $\mu(a)$           
of initial positions, but here the replacement in (\ref{eq:Q12_1}  ) of $\phi$ by $\Phi$ takes (\ref{eq:Q12_2} ) to a generalised  form involving the Fourier transform of $\mu$(a).

\subsection { General case of `$N$' walkers, `$N+1$' walls}

The generalisation of the above development to $N+1$ fixed periodically located absorbing walls $\textless$ $\textgreater$ and $N$ intervening walkers $\textgreater$$\textless$ can be treated in a similar fashion. This is true for both walkers with specific initial positions and for those with distributions of initial positions. For simplicity we give just the development for the first, simpler, case which generalises trivially.

We treat $N+1$ walls $\textless$ $\textgreater$ (denoted by $j = 1$ to $N+1$) stuck at sites   $b_j = 2(j-1)b$    and $N$ intervening walkers $\textgreater$$\textless$ , position variables $x_k$ , $k=1$ to $N$, initially at sites   $a_k$ = $b_k$+$d_k$    which are distant $d_k$  and $b-d_k$  from the walls on their left and right respectively.

It can be verified that the image procedure used above to treat the case $N=1$ again closes for this case of general $N$. However, here each walker $k$ has its own image system  arising from the two adjacent fixed walls on either side of it.  Each such system is just like that for the single-walker-between-two--walls case just discussed, and in particular is periodic with the same periodicity.  So, up to the time of the first walker annihilation, the probability of survival to time $t$ of all $N$ walkers  to positions { $x_k$ },  and of all $N+1$ walls, is the following product of factors for each walker:

\begin{equation}\label{eq:QNN_1}
  Q^{N,N+1} (\{x_k\}, \{a_k\}; \{b\})  = \prod_{k=1}^N Q^{1,2} (x_k,a_k;b_k,b_{k+1}).             
\end{equation}

Here  either of the two equivalent forms  for   $Q^{1,2}$    ( given in (\ref{eq:Q12_1}) and (\ref{eq:Q12_2}) ) can be used.

It is easily verified (using either form) that $Q^{N,N+1}$ vanishes for any $x_k$ equal to $b_k$ or $b_{k+1}$ (walker at wall). Until the first such event it provides the proper value for the survival probability distribution since all walkers are then between ``their" two walls.
The corresponding ``survival-everywhere" probability $\tilde{Q}^{N,N+1}$(\{$a_k$\};\{$b_j$\}) is obtained from (\ref{eq:QNN_1}) by integrating each $x_k$ over all possible values (between walls).

The late-time dependence of both types of survival probability for general $N$ is easily obtained from the product of the forms (\ref{eq:Q12_2})  for the $N=1$ case, which involves the same discreteness of wave vectors, related by $q_n$ = $\frac{n\pi}{b_2 - b_1}$ to the wall spacing.  This again gives  exponential decay at late times, the same  for the two probabilities,but now for general $N$ the product of $N$ factors makes the decay time 

\begin{equation}\label{eq:N_b2_b1}
 [{(b_2-b_1)/\pi}^2] / N            
\end{equation}

Concerning early times,  (\ref{eq:QNN_1})  shows using  (\ref{eq:Q12_1})   that then each survival probability is dominated by the contribution from the free-walk term in  (\ref{eq:Q12_1})  unless any walker starts near a wall, in which case the corresponding image term in (\ref{eq:Q12_1}) gives  an appreciable addition to the  free-walk term after a time of order  $d_k^2$   or  $(b - d_k)^2$ .  
The dominant early-time term, needed below, for the survival-everywhere probability for $N$ walkers starting midway
 between their walls is

\begin{equation}\label{eq:survival_prob_exp}
 [1 - \frac{1}{\sqrt{\pi t}} exp(-b^2/16 t)]^N             
\end{equation}

Among other quantities described in this paper, the analytical predictions for the survival-everywhere probability can be compared with numerical results from simulation.  Simulations were carried out using an appropriate algorithm for the no-arrow-overlap dynamics, ie with mobile walkers and immobile walls, corresponding to the process $A+B \rightarrow 0$, and also for the process  $A+A \rightarrow 0$.
After the generation of appropriate initial states simulations were run up to times long enough to exceed expected (and actual) characteristic times and finite cut-offs from finite size. Averaging over many histories ($\sim 10^5$) was exploited. For both types of process a wide range of values of $N$ and of wall separations were used.

\begin{figure}
\includegraphics[height=5.8cm,width=7.8cm, clip=true]{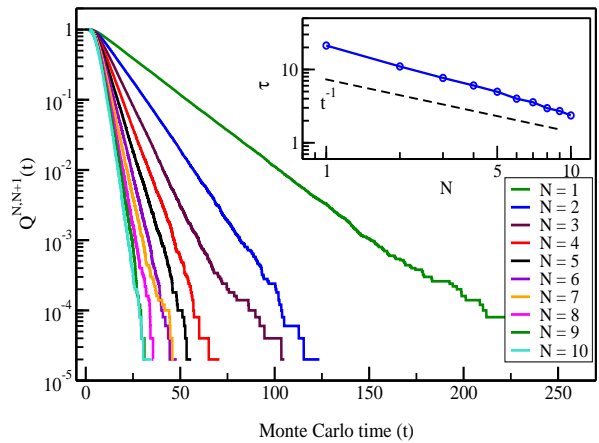}
\caption{Probability of all-walker survival for $A+B\rightarrow0$ : variation with $N$. The inset shows $\tau$ versus $N$. }
\label{fig:surv_A+B_tau_vs_N}
\end{figure}

\begin{figure}
\includegraphics[height=5.8cm,width=7.8cm, clip=true]{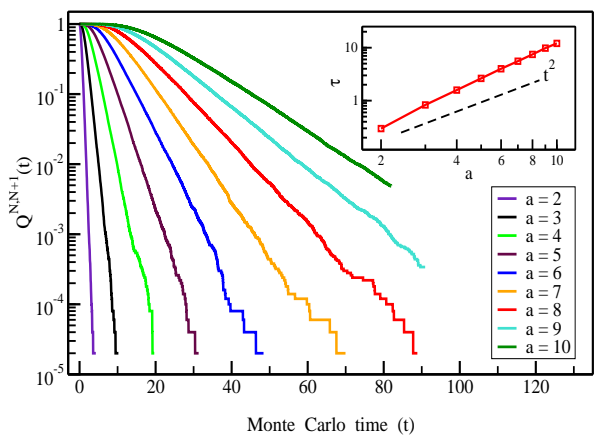}
\caption{Probability of all-walker survival for $A+B\rightarrow0$ : variation with initial spacing $a$. The inset shows $\tau$ versus $a$.}
\label{fig:surv_A+B_tau_vs_a}
\end{figure}

For the first case, with no-overlap constraint, each mobile walker was initially taken to be midway between its fixed confining walls  $(a=b/2)$.  The results for this case exhibited in figures \ref{fig:surv_A+B_tau_vs_N}  and  \ref{fig:surv_A+B_tau_vs_a} are all for the survival-everywhere probability of all $N+1$ walls and $N$ walkers. The figures \ref{fig:surv_A+B_tau_vs_N}, \ref{fig:surv_A+B_tau_vs_a}  both give clear evidence of exponential decay after a time roughly comparable to that, $(2a/ \pi)^2 /N$  , predicted by the theory.

The wide range of $N$ values covered by the results in figure \ref{fig:surv_A+B_tau_vs_N}  allow an accurate estimate of the $N$-dependence of the decay time. The inset showing log $\tau$  versus  log $N$  gives exponent 0.94. Similarly in figure \ref{fig:surv_A+B_tau_vs_a} the wide range of $a$ values used in the log-log plot in the inset allows the exponent estimate 2.22. The exponents appear to be converging towards the theoretical values (1 and 2) as the simulation runs get longer.

\begin{figure}
\includegraphics[height=5.8cm,width=7.8cm, clip=true]{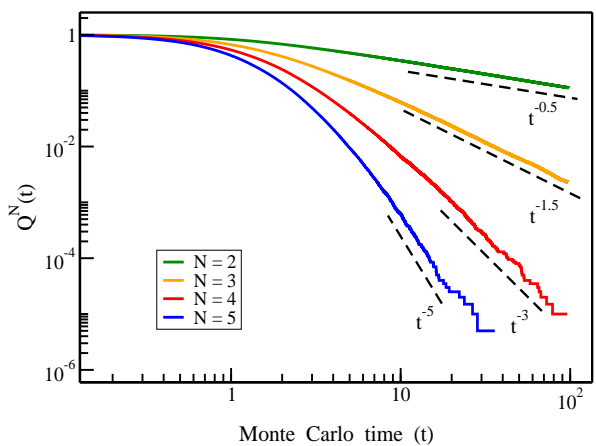}
\caption{Probability of all-walker survival for $A+A\rightarrow 0$. The dashed lines indicate the power law decays predicted in Ref. \cite{fisher}}
\label{fig:surv_A+A}
\end{figure}

Figure \ref{fig:surv_A+B_tau_vs_a} also exhibits, particularly for the largest $a$'s, the type of early-time behaviour theoretically predicted in (\ref{eq:survival_prob_exp}). The late exponential decay in figures \ref{fig:surv_A+B_tau_vs_N} and \ref{fig:surv_A+B_tau_vs_a} is in stark contrast to the behaviour in figure \ref{fig:surv_A+A} for the $A+A \rightarrow 0$ case.
 
Here the results are consistent at large times with power-law decay,  $t^{-\gamma(N)}$  where  $\gamma(N) =N(N-1)/4,$ as predicted by the ``vicious walkers" theory \cite{fisher}.

As just discussed, it has been found that for finite $N$ in  the $A+B\rightarrow0$ process (I) the probability that all walkers survive up to $t$ falls exponentially, ie much faster than  (the power law) for the $A+A\rightarrow0$ (process II).  However  for persistence (see § V-B) the fraction of walkers surviving falls slower for process I than for process II. 

Such observations raise general questions about possible effects, including crossovers, of particular conditions considered, and of such things as finite number versus finite density of walkers, etc.  

A particular condition which affects the decay rate of the all-walker-survival prob $Q^N$...is the initial gap distribution.  
With all gaps starting at the same value, $b$, Eq(\ref{eq:survival_prob_exp}) gives $Q$ decaying exponentially with rate $\sim b^{-2}$. For the same initial condition in process II  the corresponding decay starts similarly but the greater possibility of gaps increasing allows a crossover to later power-law decay 
\cite{fisher}.

For persistence the late time regime of interest is where a large fraction of the walkers have been annihilated, the $b$'s have become large, and the rare survivors are of interest. All this occurs well after the crossover just referred to, which allows a less drastic diminution of the number of surviving walls in II than in I. But simple qualitative arguments do not easily explain the  faster decay of unflipped arrowheads in II than I.
Different initial gap distributions can lead to different or no crossovers. For example distributions giving weight to very small gaps  allow much smaller characteristic and crossover times, and most initial distributions give rise to long-time distributions which are invariant except for a common inflation of gaps with time ($\sim t^{\frac{1}{2}}$).

\section {Coarsening and Persistence}

In this section, we discuss the dynamics of approach to equilibrium in large systems. Every arrowhead configuration has alternating $A$ and $B$ domain walls and the approach to steady state involves the decrease of their number under annihilation kinetics $A+B \rightarrow 0$. We study the case in which $A$ walls diffuse while $B$ walls are stationary, as appropriate to the limit $\rho \rightarrow \infty$; the effects of $B$ wall diffusion will be the subject of Section VI.

We monitor domain wall densities and persistence properties of arrowheads as the system evolves towards equilibrium. We use two types of initial conditions:
a) Random: with random placement of arrowheads, leading to random locations of domain walls.
b) Periodic: with equally-long alternating stretches of ...$\textgreater \textgreater \textgreater$... and ...$\textless \textless \textless $..., implying a periodic arrangement of domain walls.
In either case, $A$ and $B$ domain walls alternate in sequence.

As we will see, there are some similarities to, and some marked differences from, Glauber-Ising systems, whose dynamics is governed by the single-species annihilation process $A+A \rightarrow 0$.

Evidently, arrowhead kinematics leads to alternating $A$ and $B$ walls in all configurations, including initial conditions. This is very different from the situation in several studies of $A+B \rightarrow 0$ where $A$ and $B$ particles are placed at random (\cite{redner_1,toussaint,bplee,bray,geza_odor}). In the latter case, concentration fluctuations decay very slowly and dominate the late time dynamics. Our study is closer to that of \cite{dandekar,bondhyo,kwon_kim,lee} as will be discussed in Section VI.

Numerical results are obtained using Monte Carlo simulations where system size is typically $L$ = 10000 and averaging is done over few hundreds of histories. 

\subsection {Domain Wall Density}

As time passes, diffusing $A$ walls annihilate with stationary $B$’s. In Section IV with a finite number of walls, we saw that the all-walker survival probability decays exponentially in time, with a decay time which varies inversely with the number of walkers $N$. This time is vanishingly small in the present context, where the number of walls is macroscopic. We focus, to start with, on the wall density and how it decays in time.

\begin{figure}
\includegraphics[height=5.8cm, width=7.8cm, clip=true]{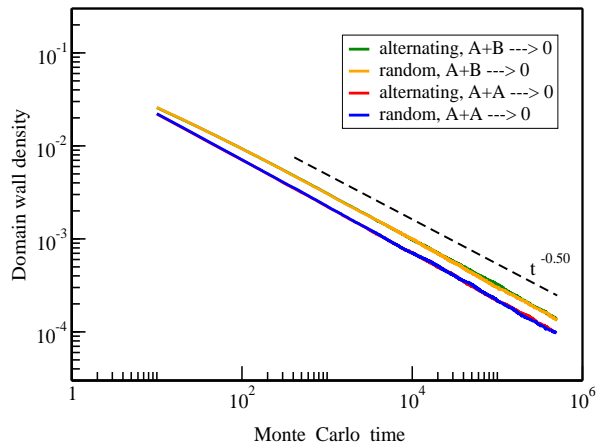}
\caption{ Decay of domain walll density} 
\label{fig:domain_wall_decay} 
\end{figure}

Since the underlying dynamics of the mobile species is purely diffusive, we expect that the result would be qualitatively similar to that for $A+A \rightarrow 0$ for the Glauber-Ising case. This is borne out by our numerical results which show the decay of wall density for the two systems, with initial conditions a) and b) for each case. The density follows a $t^{-\delta}$ decay with $\delta=\frac{1}{2}$ in all cases. See fig. \ref{fig:domain_wall_decay}

This result agrees with that of \cite{korea_1} where it was found that $\delta = \frac{1}{2}$ when the rate of the ASEP-like moves is set equal to zero.

\subsection {Persistence}

Persistence quantifies how much of particular properties of the initial configuration survive without change up to time $t$ \cite{derrida,derrida_2,satya_3}. These properties could either be local (for example the orientation of individual arrowheads) or global (for instance the majority orientation of arrowheads). 

In spin models, it is customary to monitor the fraction $F(t)$ of persistent spins, i.e. those which have not flipped up to time $t$. Equivalently, $F(t)$ can be thought of as the probability that a given spin has not flipped. The persistence probability follows a power law decay, $F(t) \sim t^{-\theta}$. A closed form expression has been obtained for $\theta$ for the $q$-state Potts models on a one-dimensional lattice; for the Ising model, the result is $\theta = \frac {3}{8}$.

\begin{figure}
\includegraphics[height=5.8cm, width=7.8cm, clip=true]{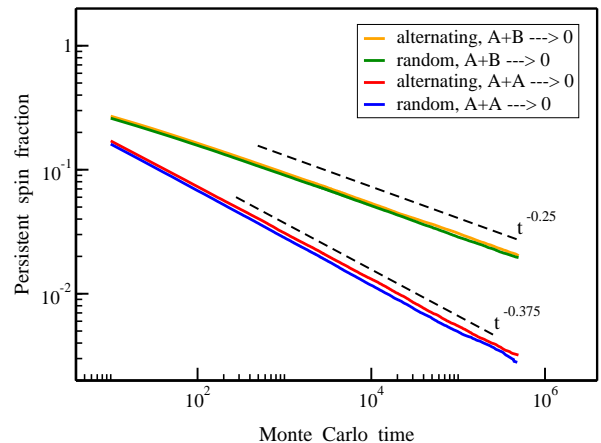}
\caption{Decay of persistence} 
\label{fig:persist} 
\end{figure}

Our data for the persistent fraction $F(t)$ in the arrowhead problem are shown in Fig. \ref{fig:persist} along-side the data for the Ising model. It is evident that the decay exponent is substantially different; we find the value for $\theta$ is 0.245, intriguingly close to $\frac {1}{4}$. Note that the value of $\theta$ is independent of the initial condition a) or b).

\begin{figure}
\includegraphics[height=5.8cm, width=7.8cm, clip=true]{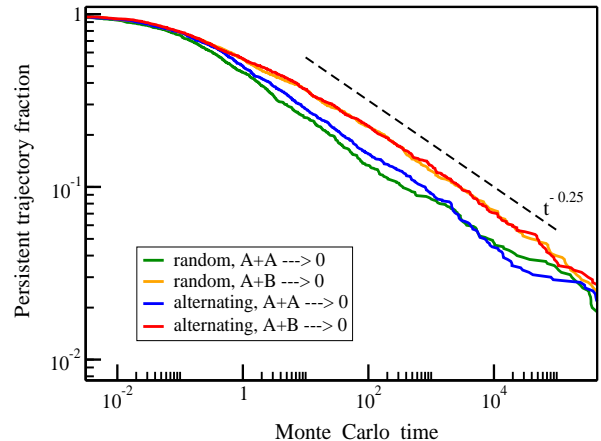}
\caption{Decay of global persistence} 
\label{fig:global_persist} 
\end{figure}

The global persistence $G(t)$ is defined as  the fraction of histories in which a global variable has not changed sign. An appropriate global variable is the `order parameter' $A(t) = A_+(t) - A_-(t)$ where  ${A_+(t) }$ and ${ A_-(t)}$ are respectively the number of right and left pointing arrowheads. From our numerical results (fig. \ref{fig:global_persist}  we find $G(t) \sim t^{-\theta_G}$ where the value of $\theta_G$ is consistent with $\frac{1}{4}$ --- which agrees with the value for $\theta_G$ in the Ising case. 

For the Ising model, the result $\theta_G = \frac{1}{4}$ was derived in \cite{satya_1} by observing that the scaled order parameter (analogous to  $A(t)/\sqrt{M}$) obeys random walk dynamics provided we redefine the time variable to be $\tau = t^2$. The only input required for this argument is that the density of surviving walkers falls as $t^{-\frac{1}{2}}$. Since this is true in our case as well, the result $\theta_G = \frac{1}{4}$ holds here as well.

 Since individual entities generally decay faster than global variables, we expect that the value of $\theta$ should be greater than or equal to $\theta_G$. Thus we surmise that the value of the site persistence exponent is also $\frac{1}{4}$. In all other systems we are aware of, $\theta \textgreater \theta_G$; this is the first instance where the equality $\theta = \theta_G$ seems to hold.

\section {Effects of `$B$' Wall Diffusion}

In order to investigate the universality of the results obtained in Section V, we have studied the effects of letting the $B$ walls diffuse, by allowing the moves of Eq. (\ref{eq:moves_2}) in addition to those of Eq. (\ref{eq:moves_1}). 

In the original arrowhead problem, movement of a $B$ wall would require a large enough gap to open up for it to occur without violating the hard-core constraint; in the equilibrium state, this would be extremely rare if the density is large. Thus the moves considered (Eq.(\ref{eq:moves_2})) are best viewed as put in ‘by hand’; they would need a softening of the hard-core constraint between arrowheads at $B$ interfaces in order to happen. Our motivation in studying such moves is purely to investigate the theoretical question of universality with respect to allowing $B$ wall diffusion, partly motivated by the importance of the question for the $A+B \rightarrow 0$ problem, and partly by the observations of \cite{korea_1} where the authors observed a continuous variation of decay exponents as some rates were changed in their model.

We first examined numerically the manner in which the density of walls decays. We found that it follows $\rho (t) \sim t^{-\delta}$  where the value of $\delta$ appears to remain fixed at $\frac{1}{2}$, as for the arrowhead and Ising cases. The data is not displayed, but falls in between the two limiting cases shown in Fig.\ref{fig:domain_wall_decay}. This is in contrast to the variation of $\delta$ observed in \cite{korea_1} when the rate of the ASEP-like move in their model was varied.

Coming to the question of persistence, namely the probability that a given spin has not flipped up to time $t$, we note that the question may be posed in the context of a finite number of walls, as considered in Section IV. In fact, the simplest case is most illuminating: three walkers, with a diffusion constant $D$ for the outer two walkers, and $D^{\prime}$ for the central walker. The survival problem can be solved exactly by a mapping to the motion of a composite particle in a wedge-shaped domain \cite{redner}. The result is 
\begin{equation}
 \theta = \frac {\pi}{2~cos^{-1} \frac {D^{\prime}}{D+D^{\prime}}}             
\end{equation}

The significant point is that $\theta$ depends explicitly on the ratio $D^{\prime} / D$. It is instructive to check limiting cases. If the central particle is a stationary $B$ particle surrounded by two $A$’s, we obtain $\theta$ = 1, implying a faster decay than with a single $A$ particle ($\theta =\frac{1}{2}$). On the other hand, if the central particle is an $A$ surrounded by two $B$ particles, we obtain $\theta=\infty$, consistent with the exponential decay found in Section IV.

\begin{figure}
\includegraphics[height=5.8cm, width=7.8cm, clip=true]{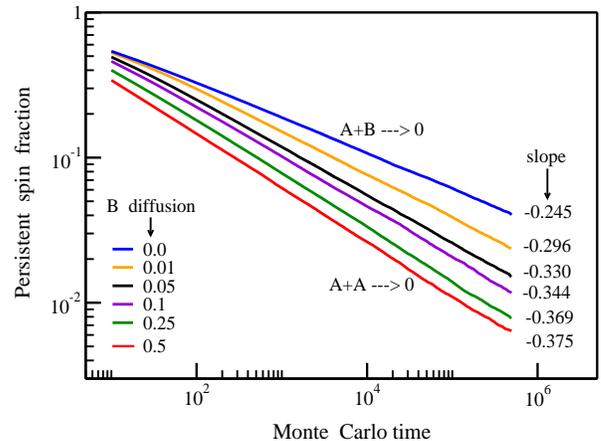}
\caption{ Variation of power laws characterizing decay of persistence on varying diffusion constant for B particles } 
\label{fig:persist_var} 
\end{figure}

To see whether this dependence of the power law exponent on diffusion constants stays also for persistence properties in a thermodynamically large system, we numerically studied the fraction of persistent spins $F(t)$ as a function of $t$, and found a definite variation of the power law exponent $\theta$ as the ratio of diffusion constants $u^{\prime}/u$ is varied. The data Fig. \ref{fig:persist_var} indicate a smooth variation of $\theta$ from about 0.25 in the arrowhead model $(u^{\prime}/u=0)$ to 0.375 in the Ising model $(u^{\prime}/u=1)$. 

However, the global persistence $G(t)$ continues to show a $t^{-\frac{1}{4}}$ decay as in the arrowhead and Ising cases. This is not unexpected, as following the argument in Section V, the decay of $G(t)$ is related to the manner in which the density of domain walls decays, and as we have seen above, this does not depend on  $u^{\prime}/u$.

\section{Conclusion}

In this paper we have studied a system of stochastically reorienting arrowheads in one dimension. Our study of the static properties and the dynamics of approach to the steady state has touched on several issues, and we discuss our results in that context.
The equilibrium state of our system approaches an orientationally ordered state as the density increases. The origin of order is entropic, as follows from the fact that arrowheads may never overlap, implying that every allowed configuration is equally likely. For a fixed total length, it is evident that the entropy of translation is largest in an orientationally ordered state, as then there is no constraint on the locations of arrowhead vertices other than that they must maintain a sequence. This contribution to the entropy dominates at large density; at moderate values of $\rho$, it must compete with the configurational entropy of the locations of interfaces. The spatial extent of the order is quantified by the correlation length and the persistence length, and the calculations in Section III show that both diverge as $e^{\rho}$ as $\rho \rightarrow \infty$.

Understanding the dynamics of approach to the ordered state is greatly facilitated by the observation that there are two types, $A$ ($\textgreater \textless$) and $B$($\textless \textgreater$), of arrowhead interfaces obeying $A+B \rightarrow 0$ dynamics. Our study has a bearing on a couple of interconnected issues in this two-species annihilation problem. In this problem, the influence of initial conditions on long-time decays has long been recognized. Initial conditions in which the sequence of $A$’s and $B$’s is random give rise to long-lived concentration fluctuations in the number of $A$ and $B$ particles, which in turn give rise to multiple length scales and slow down the dynamics strongly \cite{privman_book}. By contrast, under initial conditions in which  the imbalance between $A$ and $B$ particle numbers remains of the order of unity in every stretch, the two species are well mixed, and the decay exponent $\delta = \frac{1}{2}$. Recent work \cite{dandekar} has shown that when alternation follows the pattern $A^nB^nA^nB^n ...$ there are multiplicative logarithms for even $n$. Such initial conditions have been studied in the context of conserved lattice gas models, on identifying pairs of sites with $A$ and $B$ particles \cite{bondhyo,kwon_kim,lee}. In the arrowhead model, even a random placement of arrowheads results in strict alternation of $A$ and $B$ interfaces; thus our result $\delta = \frac{1}{2}$ (section V) with no evidence of logarithms is fully consistent with the above.
Another important issue concerns the fact that in our problem, the diffusion constants for $A$ and $B$ particles are not equal; in fact the $B$ particles do not diffuse at all in the limit of infinite arrowhead density. How pertinent is this for long time decays of survival probabilities under $A+B \rightarrow 0$ dynamics? With a finite number $N$ of walkers, we showed that the all-walker survival probability falls exponentially rapidly in time, with the decay time being inversely proportional to $N$. This contrasts strongly with the power law decay with power $N(N-1)/4$ found with equal diffusion constants, as appropriate to $A+A \rightarrow 0$. In the thermodynamically large system, we found the persistence probability falls as a power law for both $A+B \rightarrow 0$ and $A+A \rightarrow 0$, but importantly, the powers differ, being close to $\frac{1}{4}$ in the former case, and $\frac{3}{8}$ in the latter.

We conclude with some comments on universality. The studies of \cite{korea_1} on a generalized model with ASEP-like moves together with annihilation indicate a violation of universality, for example  in the power laws for domain wall density, as the ASEP rates are changed. In the reorienting arrowhead model, there is no ASEP move. We find that on varying the diffusion constant for $B$ particles from 0 to the value for $A$ particles, globally averaged properties such as the number of domain walls, or global persistence, remain universal. But more delicate properties such as single site persistence are found to exhibit a continuous variation of power laws as the ratio of diffusion constants is varied. This indicates a violation of universality, but at a weaker level than in the studies of \cite{korea_1}.

\section {Acknowledgements}
MDK acknowledges Institute of Mathematical Sciences (IMSC), Chennai, India for selection in associateship programme for college teachers. MB acknowledges the hospitality of the Rudolf Peierls Centre for Theoretical Physics at the University of Oxford where part of this work was done and the award of the J C Bose Fellowship by the Department of Science and Technology, India. He thanks Satya Majumdar for a useful discussion.

\end{document}